\input aa.cmm
\voffset=1truecm
\def\etal{et al.}
\newcount\fignumber\fignumber=1
\def\nfig{\global\advance\fignumber by 1}
\def\fignam#1{\xdef#1{\the\fignumber}}
\def\nompsy#1{}
\fignam{\FROTCUR}   \nfig
\fignam{\FEVOLRONE} \nfig
\fignam{\FEVOLRTWO} \nfig
\fignam{\FEVOLRTHR} \nfig
\fignam{\FPROFIL}   \nfig
\fignam{\FVELLOST}  \nfig
\fignam{\FVELLOSC}  \nfig
\fignam{\FSCHEMA}   \nfig

\MAINTITLE{Counter-rotating bars within bars}
\AUTHOR{D.~Friedli}
\INSTITUTE{Observatoire de Gen\`eve, CH-1290 Sauverny, Switzerland}
\OFFPRINTS{D.~Friedli, fri@scsun.unige.ch}
\DATE{Received 23 November 1995 / Accepted 2 February 1996}

\ABSTRACT{
The dynamical stability and evolution of disc galaxies with different
disc thickness as well as various fraction and concentration of
stellar counter-rotation is investigated with self-consistent
numerical simulations. In particular, systems of nested,
counter-rotating, stable bars are presented. As for the direct case,
the nuclear secondary bar rotates faster than the primary
bar. Contrary to the direct case, the presence of significant amount
of gas is not necessary to produce such system. Observed bars within
bars in lenticular (i.e. gas-poor) galaxies could thus be explained by
counter-rotating bars. This hypothesis can easily be tested since the
retrograde case produces distinct kinematical signatures from the
direct one.  Moreover, the best morphological similarities between
numerical models and typical SB0's are found in the models with
significant amount of stellar nuclear counter-rotation.  }

\THESAURUS{11(11.05.2; 11.11.1; 11.14.1; 11.16.2; 11.19.6)}
\maketitle


\titlea{Introduction}
Among all the eccentric species observed in the zoo of galaxies, the
``bar-within-bar'' one (i.e. a secondary nuclear bar embedded in a
primary bar) is strange enough to draw astronomer's attention (e.g. de
Vaucouleurs 1974, 1975; Buta \& Crocker 1993; Friedli \& Martinet
1993; Shaw \etal\ 1993, 1995; Wozniak \etal\ 1995; Friedli 1996;
Friedli \etal\ 1996). In particular, these systems have been proposed
as a possible mechanism for gas fueling in active galactic nuclei
(Shlosman \etal\ 1989, 1990).  Numerous galaxies with misaligned
nested stellar bars have been observed over the years by different
authors in barred spirals or lenticulars suggesting a healthy
population.

Various explanations can be given for the existence of such shapes: i)
Projection effect of two perpendicular stellar bars.  This cannot
explain the small angles between the two bars observed in some almost
face-on galaxies. ii) Real misalignment of two stellar bars with the
same pattern speed. A very short lifetime is expected since strong
gravitational torques will quickly align the two bars. iii) Action of
a massive central leading gaseous bar onto stars (Shaw \etal\ 1993).
This cannot explain the existence of both leading and trailing
secondary bars (Buta \& Crocker 1993; Wozniak \etal\ 1995). iv) Real
misalignment of two stellar bars with two {\it different} pattern
speeds but the {\it same} direction of rotation. This nicely solves
all the above problems. Moreover, Friedli \& Martinet (1993; see also
Combes 1994) have demonstrated with their self-consistent 3D N-body
simulations with gas and stars that such systems can exist and be
stable over many bar rotations. In this scheme, the significant
primary bar-driven gas fueling plays a key role to trigger a dynamical
decoupling of the central regions which start to rotate faster and
independently. v) Real misalignment of two stellar bars with two {\it
different} pattern speeds but the {\it opposite} direction of rotation
as already simulated in 2D by Davies \& Hunter (1995, 1996) and in 3D
by Friedli (1996).  Contrary to explanations iii) or iv), large
amounts of gas are not needed in this case.  If a secondary bar can
easily rotate in the opposite way in comparison with a primary bar in
a purely collisionless system, this mechanism could then be relevant
for the gas-poor lenticular galaxies.  I thus suggest that the
secondary bar could counter-rotate in some (if not all) of these
galaxies.

Below are presented various detailed numerical simulations of the
dynamical evolution of disc galaxies with various fraction and
concentration of stars in counter-rotation. Of particular interest are
the models of nested, counter-rotating, stable bars.  Complete
formation models of such galaxies are deferred to a subsequent paper;
the aim here is only to demonstrate that such systems can exist, be
long-lived, and to have some insight on how they look.  At first
sight, this could appear as a purely academic exercise but the
discovery of galaxies with massive counter-rotating stellar discs
(NGC~4550, Rubin \etal\ 1992, Rix \etal\ 1992; NGC~7217, Kuijken 1993,
Merrifield \& Kuijken 1994; NGC~3593, Bertola \etal\ 1996), or gaseous
discs (NGC~3926, Ciri \etal\ 1995; see Galletta 1996 for a recent
review) raises also the general interesting question of retrograde
dynamics (e.g. Levison \etal\ 1990; Sellwood \& Merritt 1994).
Moreover, a large fraction of luminous elliptical galaxies contain
kinematically decoupled, counter-rotating, peculiar cores as well (see
e.g. Bender 1995).  Anyway, satellites on retrograde orbits exist and
by dynamical friction some of them should have been swallowed by their
host galaxies in the past, and have thus brought fresh retrograde
angular momentum into the disc system.  The formation of massive
counter-rotating disks in spiral galaxies has been recently
investigated by Thakar \& Ryden (1995).

This paper is constructed as follows: The code, the models and the
definitions used are given in Sect.~2.  Section~3 briefly recalls the
results obtained in the direct cases whereas Sect.~4 presents in a
detailed way the formation, evolution, morphology and kinematics of
the retrograde cases. A general discussion as well as possible
formation scenarios are presented in Sect.~5 and my conclusions are
summarized in Sect.~6.

\titlea{Code, models and definitions}
\titleb{The code}
Our self-consistent 3D simulations of disc galaxies are modeled with a
purely collisionless component (stars) as well as with a dissipative
one (gas) in some models.  The complete code description is given in
Pfenniger \& Friedli (1993) for the gravitational part (Particle-Mesh
method), and in Friedli \& Benz (1993) for the hydrodynamical part
(Smooth Particle Hydrodynamics technique).  Typically, the number of
star particles is $2 \cdot 10^5$ and the number of gas particles, when
present, is $1 \cdot 10^4$. The time integration is performed with a
leap-frog algorithm (constant time-step $\Delta t\!=\!0.25$) in purely
collisionless systems and with a Runge-Kutta algorithm (adaptative
time-step) in mixed systems with gas and stars.

\titleb{Models}
Initially, the models are axisymmetric. Their stellar parts are
distributed according to a superposition of two Miyamoto-Nagai models
of mass $M_1$ and $M_2$, with vertical scale-heights $b_1$ and $b_2$,
and horizontal scales $a_1 + b_1$, $a_2 + b_2$; the mass of the
gaseous disc is $M_g$ (for details see Friedli \& Benz 1993). The
rotation velocity and the three components of the velocity dispersion
are numerically computed on the grid so as to satisfy the equilibrium
solutions of the stellar hydrodynamical equations.  The systematic
rotation velocity of the component 1 can be put either like the one of
the component 2 (direct models $D_i$) or in the opposite direction
(retrograde models $R_i$).  Obviously, this does not correspond to
a very realistic formation scheme, but the aim here is mainly to prove
that such configurations can exist and be stable.  A discussion
concerning complete formation scenarios is given in Sect.~5.

\titleb{Units and definitions}
For computational convenience, the calculations are performed in units
suitable for galactic problems. The unit length is chosen to be the
kpc, the unit mass represents $2 \cdot 10^{11} \, \rm M_\odot$, and
with $G\!=\!1$ the time unit is $1.05$ Myr.

The bar with the smaller extent will be referred as the ``secondary
bar'' and all the quantities related with it will have the subscript
$s$.  The bar with the larger extent will be referred as the ``primary
bar'' and all the quantities related with it will have the subscript
$p$. Thus, for instance the primary and secondary bar maximum
ellipticities are respectively denoted as $e_p^{\rm max}$ and
$e_s^{\rm max}$. These values are determined with the ellipse fitting
technique (for details see Wozniak \etal\ 1995).  The angle between
the two bars is $\theta$.  Positive values are for leading secondary
bars whereas negative values are for trailing secondary bars (relative
to the primary bar rotation). The other quantities are:
\item{--}
The ratio $\alpha \equiv \Omega_s /\Omega_p$, where the $\Omega_i$ are
the bar pattern speeds.
\item{--}
The ratio $\beta \equiv l_p/l_s$, where the $l_i$ are the bar lengths.
\item{--}
The ratio $\gamma \equiv m_p/m_s$, where $m_s$ is the total mass
between the centre and $l_s$, and $m_p$ is the total mass between the
centre and $l_p$. 

\titlea{The direct models}
\titleb{Fast nested bars within bars
($\beta > 1$, $\alpha > 1$)}
This direct case has extensively been discussed and described in
Friedli \& Martinet (1993) so that only a brief summary is given here.
The basic model $D_1$ has the following initial parameters:
$M_g\!=\!0.1$, $M_1\!=\!0.1$, $M_2\!=\!0.8$, $a_1\!=\!b_1\!=\!0.2$,
$a_2\!=\!6.0$, and $b_2\!=\!0.75$.  The initial stellar rotation curve
is given in Fig.~{\FROTCUR a}.  

Schematically, the general evolution of the simulation is: 1) Slow and
simultaneous formation of the primary and the secondary bar
(i.e. $\beta \!>\!1$) . 2) Phase of two bars with different pattern
speeds (i.e. $\alpha \!>\! 1$) due to a decoupling of the central
dynamics from the outer parts.  Progressive formation of a nuclear
gaseous ring near the corotation radius CR$_s$.  3) Dissolution of the
secondary bar, the primary bar remains with the nuclear ring
anti-aligned with it.  The central gaseous mass accreted is not
sufficient to dissolve the primary bar as well.  The appearance of a
double-barred system and a nuclear gaseous ring are favoured by the
following characteristics: 1) The two stellar discs must obviously be
bar unstable and the primary bar must have an inner Lindblad resonance
(ILR) allowing for the presence of the anti-bar $x_2$ orbit family
which is essential for increasing the efficiency of the decoupling.
2) To maintain the dynamical decoupling\fonote{The dynamical {\it
decoupling} could be related to non-linear mode {\it coupling} (Tagger
\etal\ 1987)}, a significant dissipative component has to be present.
Then the material, which accumulates onto the central galaxy part,
reinforces the secondary bar strength, and prevents the fast slow down
of its pattern speed.  3) Systems with $\alpha \gg 1$ minimize the
interaction between the two bars and those with the CR$_s$ coincident
with the ILR$_p$ minimize the generation of chaos.  4) Slow enough
bar growth(s) and not too strong bar strength(s) are favourable for
the formation of a nuclear gaseous ring during the two bar phase. This
indeed insures that some gas remains outside the secondary bar region.

At $t\!=\!1200$, this double-barred model has the following
characteristics: $\alpha \approx 3.1$, the positions of the
axisymmetric ILR$_p$ and CR$_s$ are about the same, $\theta \approx
90\degr$, $\beta \approx 5.4$, $\gamma \approx 2.9$, $e_s^{\rm max}
\approx 0.39$, and $e_p^{\rm max} \approx 0.27$. Different values have
also been obtained in new recent simulations (Friedli \etal\ 1996).
The stellar rotation curve is shown in Fig.~{\FROTCUR b}.

\titleb{Leading nested bars within bars
($\beta \!>\! 1$, $\alpha \!=\! 1$, $+10\degr \la \theta \la
+80\degr$)} Shaw et al. (1993) have presented models with gas and
stars which produce two misaligned stellar bars rotating at the same
pattern speed.  First the star-gas disc is unstable and quickly forms
a primary bar with two ILR's. The gas clouds should follow the $x_2$
family but due to dissipative collisions they are in fact gradually
shifted from parallel to perpendicular orbits relative to the primary
bar and the gas settles in a {\it leading} phase-shifted bar.  If the
gas fraction is high, its gravitational influence is then sufficient
to modify the stellar component itself and to form the secondary
stellar bar. Combes (1994) indicates $\theta \approx +30\degr$ but in
principle angles $+10\degr
\la \theta \la +80\degr$ are possible depending on the model
properties, in particular its viscosity.  Unfortunately, this model
cannot account for the existence of observed {\it trailing} secondary
bars.

It is not yet clear what allows some galaxies to have one single
pattern speed and other to have two different pattern speeds.  In both
cases, the gas-star coupling is essential.  Simulations by Combes
(1994) seem to indicate that the gas viscosity plays the determining
role: Systems with high viscosity tend to decouple and form two
independent bars whereas systems with low viscosity preferentially
form a phase-shifted secondary bar.

\begfig 12 cm
\nompsy{frot.ps}
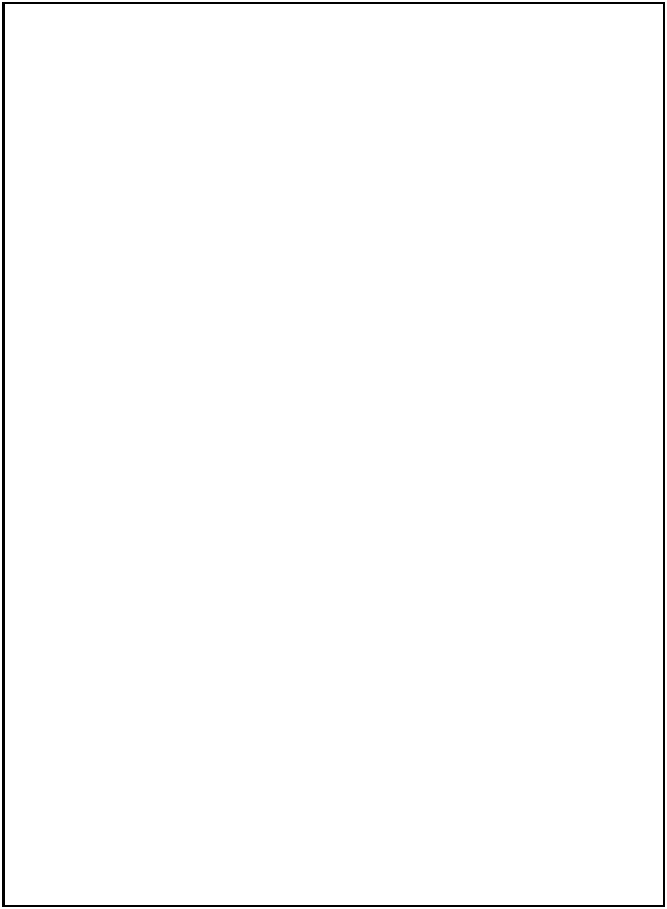
\figure{\FROTCUR a and b}{
Systematic, azimuthally averaged, tangential velocity $V_t$ as a
function of radius for models $D_1$ (dashed-dotted curve), $R_1$
(dotted curve), $R_2$ (dashed curve), and $R_3$ (solid curve).  {\bf
a} $t\!=\!0$, and {\bf b} $t\!=\!2000$ for the retrograde models and
$t\!=\!1200$ for the direct model }
\endfig

\titlea{The retrograde models}

\begfigwid 18.8 cm
\nompsy{vide.ps}
\figure{\FEVOLRONE}{
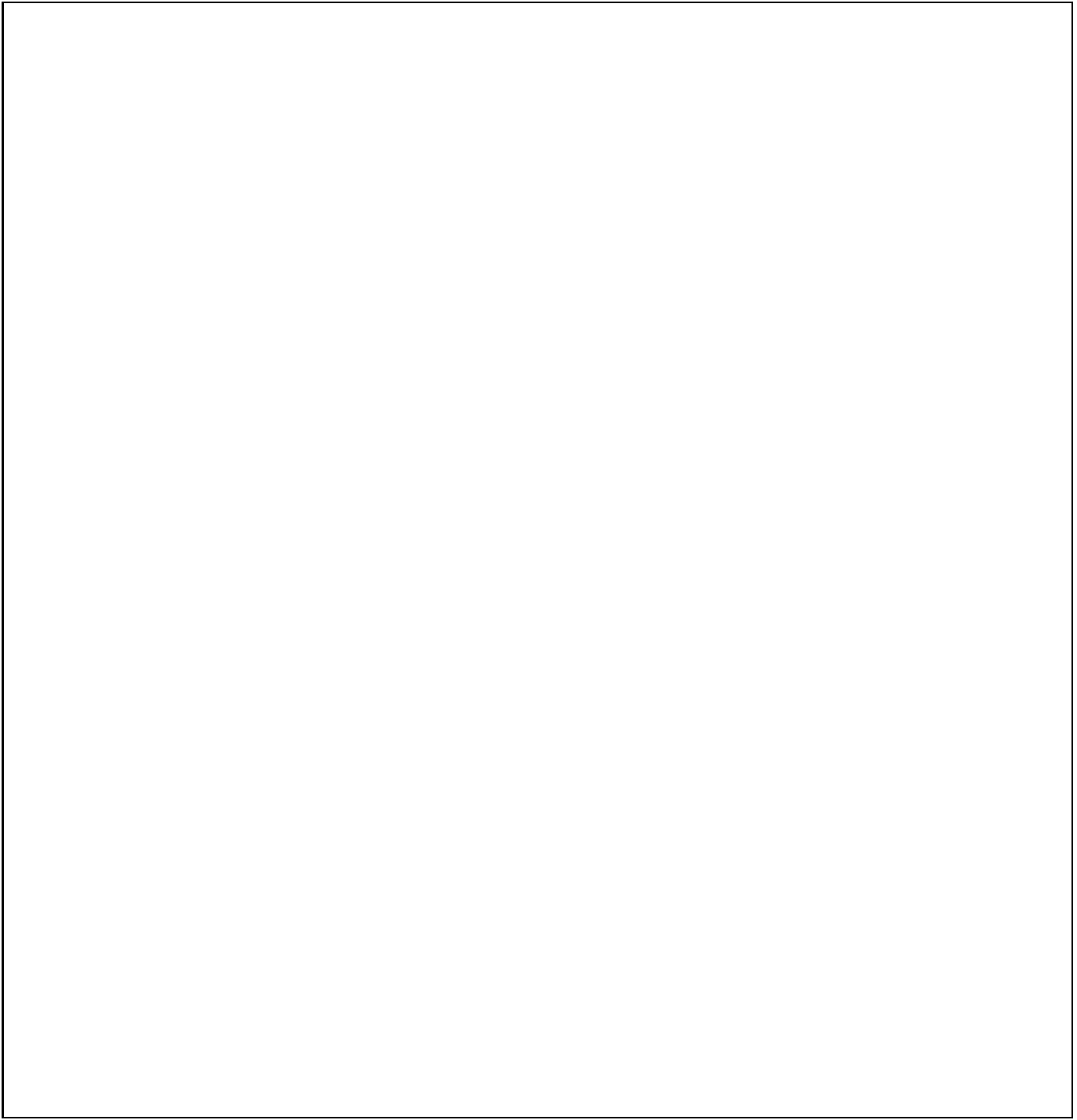
Time evolution of the stellar surface density projected in the three
principal planes of the model $R_1$ for the retrograde component with
clockwise rotation (left column), the direct component with
counter-clockwise rotation (middle column), and the total (right
column).  The side of the square frames is 10~kpc for the left column
and 20~kpc for the other ones.  The isodensity contours are separated
by 0.3~dex.  The time $t$ is indicated at the bottom left of each
frame.  The rotation of the {\it two} bars is counter-clockwise with a
rotation period of approximately 320~Myr.  Note the box-peanut shape
at $t\!=\!3000$ }
\endfig

\begfigwid 20.8 cm
\nompsy{vide.ps}
\figure{\FEVOLRTWO}{
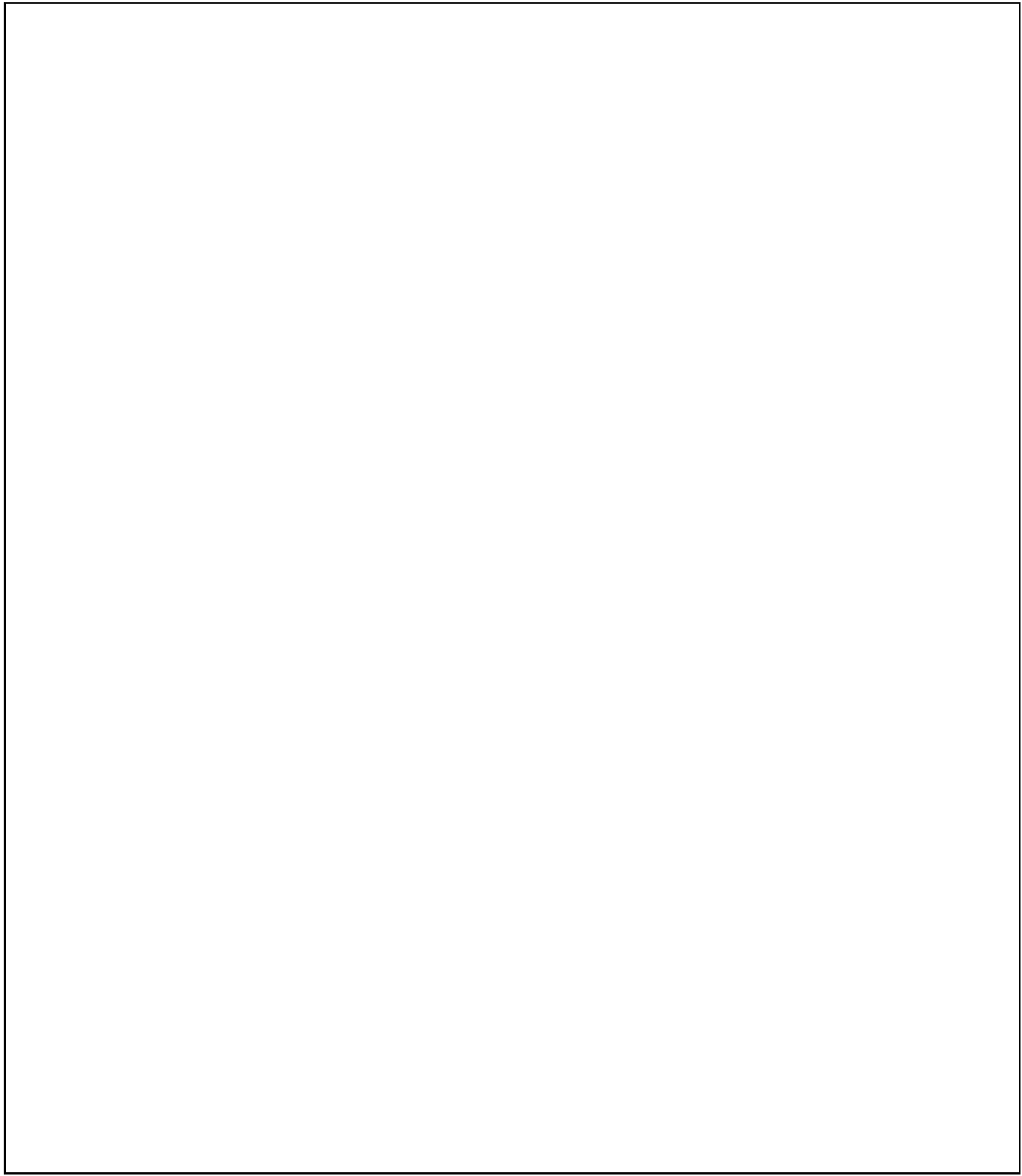
The same as Fig.~{\FEVOLRONE} but for the model $R_2$.  The retrograde
bar in the left column rotates clockwise whereas the direct bar
rotation in the middle column is counter-clockwise.  Both rotation
periods are approximately 350~Myr }
\endfig

\begfigwid 20.8 cm
\nompsy{vide.ps}
\figure{\FEVOLRTHR}{
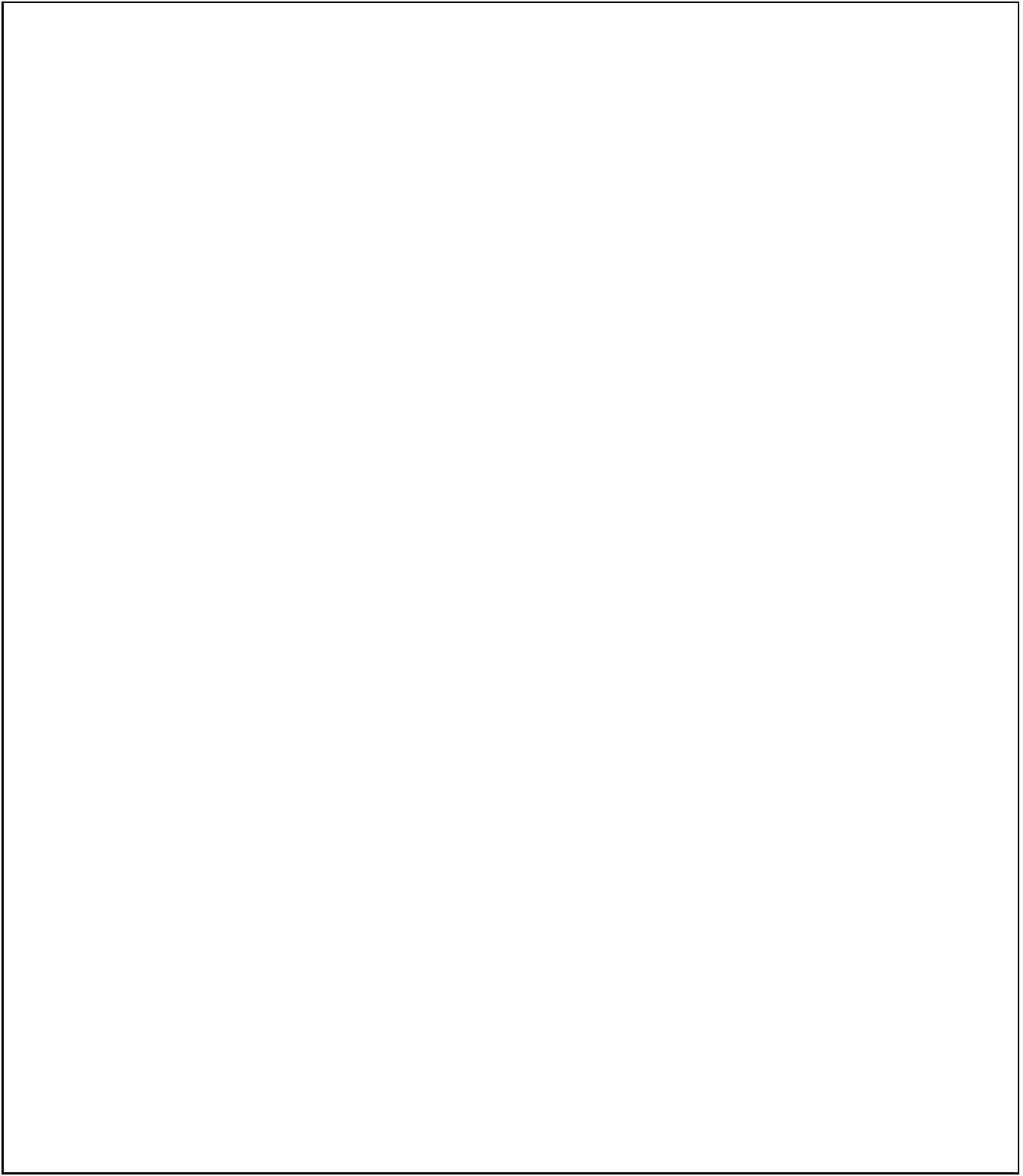
The same as Fig.~{\FEVOLRONE} but for the model $R_3$ at the beginning
of the simulation, during a fraction of the two counter-rotating bar
phase, and at the end of the simulation.  The rotation periods are
approximately 240~Myr for the large-scale primary bar
(counter-clockwise rotation) and 130~Myr for the small-scale secondary
bar (clockwise rotation) }
\endfig

\titleb{Nearly perpendicular nested bars within bars 
($\beta \!>\! 1$, $\alpha \!=\! 1$, $\theta \approx -80\degr$)} The
model $R_1$ has the following initial parameters: $M_g\!=\!0.0$,
$M_1\!=\!0.2$, $M_2\!=\!0.8$, $a_1\!=\!1.0$, $a_2\!=\!6.0$,
$b_1\!=\!b_2\!=\!0.75$.  The retrograde component represents 25\% of
the mass of the direct component but only about 5.7\% of the angular
momentum (in absolute value).  The counter-rotating component is not
strongly centrally concentrated and the systematic tangential velocity
$V_t$ is only slightly negative near the centre (see Fig.~{\FROTCUR
a}).

The time evolution of this model is presented in Fig.~{\FEVOLRONE}.  A
prominent bar quickly develops in the direct disc. This primary bar
(sustained by the $x_1$ periodic orbit family) totally dominates the
dynamics so that the retrograde disc is first essentially influenced
by the $x_4$ 2D periodic orbit family, and then around $t\!=\!2000$ by
the anomalous 3D periodic orbit family which bifurcates from the $x_4$
(1/1 vertical resonance). These families are slightly anti-bar so that
the retrograde disc becomes weakly barred. This bar is always nearly
perpendicular to the primary bar but remains almost indistinguishable
when the total surface density is plotted (nearly round centre).  The
bar looks very nicely like typical SB0's, i.e. the isodensity contours
are round near the centre, then take the shape of an elongated
hexagon, and finally become close to rectangular.  

Both bars have the same pattern speed ($\approx 20 \, \rm km \, s^{-1}
\, kpc^{-1}$).  Thus, {\it the secondary bar pattern rotates in
the opposite direction in comparison with the stars which compose it!}
This system lasts about $2 \, \rm Gyr$.  At $t\!=\!3000$ about 58\% of
the angular momentum of the retrograde component has however been
transferred.  As a consequence, the retrograde component becomes
nearly spherical and $V_t$ is no more negative near the centre (see
Fig.~{\FROTCUR b}).  At $t\!=\!2000$, this double-barred model has the
following characteristics (total surface density): $\alpha \approx
1.0$, $\theta \approx -75\degr$, $\beta \approx 16.4$, $\gamma \approx
12.5$, $e_s^{\rm max} \approx 0.25$, and $e_p^{\rm max} \approx 0.58$.
This corresponds to the time where the secondary bar is the strongest.
By comparison, $e_s^{\rm max} \approx 0.12$ at $t\!=\!1500$ and
$e_s^{\rm max} \approx 0.10$ at $t\!=\!3000$ whereas $e_p^{\rm max}$
keeps about the same value. The secondary bar length $l_s$ is also
decreasing from 0.85~kpc at $t\!=\!1500$ to 0.49~kpc at $t\!=\!3000$.

The same model but with low initial disc thickness is strongly
unstable with respect to central bar and bending modes.  Then, after
the quick dissolution of the counter-rotating component (due to the
cancelation of its retrograde angular momentum approximately twice
faster than in model $R_1$), this model evolves like model $R_1$.

\titleb{Slow large-scale bars within bars
($\beta \!=\! 1$, $\alpha \!=\! -1$)}
The model $R_2$ has the following initial parameters: $M_g\!=\!0.0$,
$M_1\!=\!0.5$, $M_2\!=\!0.5$, $a_1\!=\!a_2\!=\!6.0$,
$b_1\!=\!b_2\!=\!0.25$.  So, this corresponds to two identical thin
discs which rotate in the opposite way, very similar to what can be
observed in the edge-on galaxy NGC~4550 (Rubin \etal\ 1992; Rix \etal\
1992).  The total angular momentum is zero as well as the systematic
tangential velocity (see Fig.~{\FROTCUR a}).

As studied by Sellwood \& Merritt (1994), this kind of very flat,
counter-rotating, disc system is strongly unstable with respect to
bending modes. The same authors (see also Levison \etal\ 1990)
reported in some cases {\it the formation of two bars with the same
extent which were rotating in the opposite direction}. This is also
what I observe for this model from $t\!=\!2000$ on (see
Fig.~{\FEVOLRTWO}): Two slowly ($\approx 18 \, \rm km \, s^{-1} \,
kpc^{-1}$) counter-rotating weak bars which persist for at least $2 \,
\rm Gyr$. Both discs have also a strong $m\!=\!1$ mode.  To my
knowledge no face-on galaxies of this type have so far been observed
in nature.  This is not very surprising since the total face-on
surface density does not present very specific characteristics and
such systems could easily be missed without kinematic data.  High
initial disc thickness suppresses bending modes and counter-rotating
bars.

At $t\!=\!2000$, this double-barred model has the following
characteristics: $\alpha \approx -1.0$ so that the position of
resonances are identical, $\theta \approx +60\degr$, $\beta \approx
1.0$, $\gamma \approx 1.0$, and $e_s^{\rm max} \approx e_p^{\rm max}
\approx 0.38$.  No systematic velocity develops (see Fig.~{\FROTCUR
b}). The total angular momentum remains zero but those of each disc
decreases by about 15\%.

\titleb{Fast nested bars within bars
($\beta \!>\! 1$, $\alpha \!<\! -1$)}
The typical model $R_3$ has the following initial parameters:
$M_g\!=\!0.0$, $M_1\!=\!0.2$, $M_2\!=\!0.8$, $a_1\!=\!b_1\!=\!0.4$,
$a_2\!=\!6.0$, and $b_2\!=\!0.75$.  The idea is to explore the
viability of system with $\alpha \!<\! -1$.  The counter-rotating
component is strongly concentrated to prevent the direct component
from dominating the central dynamics as in model $R_1$.  Its mass
represents 25\% of the mass of the direct component but only about
2.9\% of the angular momentum (in absolute value).  The systematic
tangential velocity becomes strongly negative near the centre (see
Fig.~{\FROTCUR a}).  This system is similar to what can be observed in
NGC~3593 (Bertola \etal\ 1996).

With these specific initial conditions, the two discs remain
axisymmetric for 1~Gyr and then {\it each disc forms its own bar which
rotate in the opposite direction, i.e. each bar rotates in the same
way as the stars which compose it} (see Fig.~{\FEVOLRTHR}).  The
secondary bar forms first (around $t\!=\!1200$) and is faster.  The
existence of this central dynamical decoupling is made easier by the
presence of the $x_4$ orbits.  The lifetime of the counter-rotating
nested bar phase is about 1~Gyr, i.e. approximately between
$t\!=\!1500$ and $t\!=\!2500$.  Just after $t\!=\!2500$, the secondary
bar reverses its direction of rotation, overtakes the primary bar,
oscillates for some time around it and finally aligns itself with it.
At $t\!=\!3000$, the secondary bar is nearly fully dissolved into a
round centre.  This evolution is mainly driven by the significant
angular momentum transfer between the secondary and the primary
bar. Indeed, at $t\!=\!2000$ the angular momentum of the retrograde
component has already been decreased by more than 27\% and then
approximately three more percents are annihilated every 100~Myr.

Davies \& Hunter (1995, 1996) have also studied the effects which
result from putting retrograde angular momentum in galactic discs.  In
particular, they have presented models of counter-rotating bars within
bars. They start with a single 2D disc model whose central systematic
velocities are abruptly reversed.  They have essentially found the
same evolutionary behaviour as in my models.  Zang \& Hohl (1978) have
shown that a large fraction of stars on retrograde orbits strongly
inhibits the formation of the primary bar.  Here, the growth of the
primary bar is slow but it finally appears.  It is not yet clear what
fraction would be necessary to fully prevent the primary bar to form
and if stars on retrograde orbits could easily dissolve an already
formed strong bar.

The same model but with low initial disc thickness is strongly
unstable with respect to central bar and bending modes; the retrograde
angular momentum is very quickly cancelled out (5 times faster than in
model $R_3$) giving rise to a one direct bar model.

\begfig 12 cm
\nompsy{fprof.ps}
\figure{\FPROFIL}{
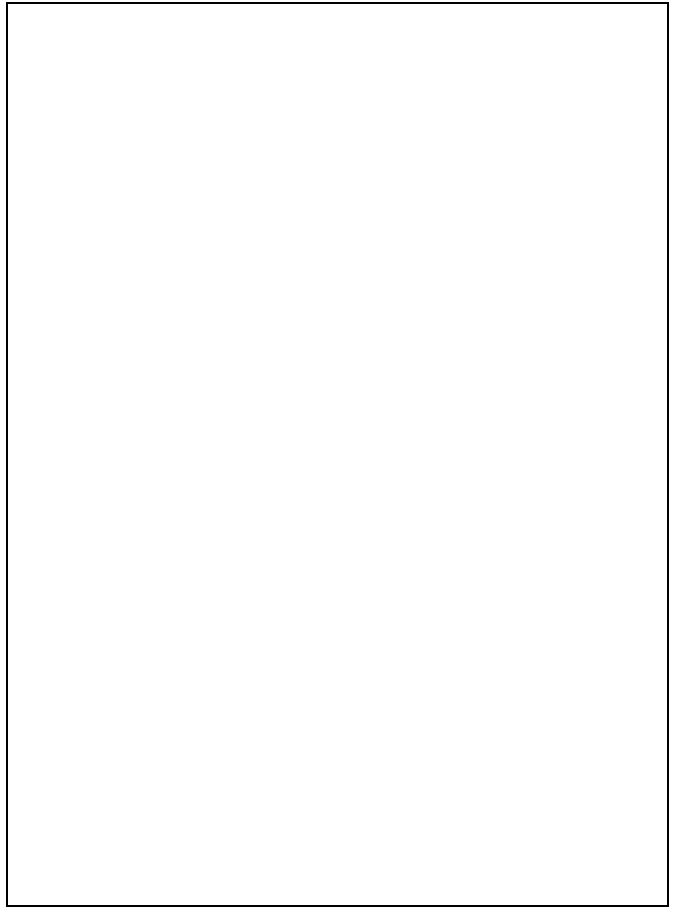
Plots of stellar surface density $\mu$, ellipticity $e$ and
position-angle PA profiles as a function of the semi-major axis of the
fitted ellipse, for the total density (solid curve), the direct
component only (dotted curve), and the retrograde component only up to
5~kpc (dashed curve).  Model $R_3$ seen face-on at $t\!=\!2000$ }
\endfig

\titlec{Morphology and resonances}
The morphological evolution of this model during a fraction of the two
bar phase is shown in Fig.~{\FEVOLRTHR} for each component as well as
for the addition of them. The ellipse fits at $t\!=\!2000$ are given
in Fig.~{\FPROFIL}. The total surface density features of the model
are $\theta \approx -60\degr$, $\beta \approx 4.1$, $\gamma \approx
2.4$, $e_s^{\rm max} \approx 0.29$, and $e_p^{\rm max} \approx 0.48$.
These values are contained between the observed intervals (Wozniak
\etal\ 1995; Friedli \etal\ 1996).  The two bars have two distinct and
nearly exponential profiles whose scale-length ratio is close to 0.13.
Very similar bar-within-bar morphology as the one of model $R_3$ have
been observed in many SB0's (e.g. NGC~1291, NGC~1543). One can thus
reasonably consider that some of these galaxies could host
counter-rotating bars within bars.

The primary bar pattern speed is $\approx 29 \, \rm km \, s^{-1} \,
kpc^{-1}$ whereas the secondary bar one is $\approx -40 \, \rm km \,
s^{-1} \, kpc^{-1}$ giving $\alpha \approx -1.4$. So, the secondary
bar does not rotate that fast, only about half the value of the direct
case although both cases have similar $\beta$ and $\gamma$.  As a
consequence and contrary to the direct case, the secondary bar does
not end close to its corotation radius but near its outer ILR radius
(cf. Fig.~{\FPROFIL}). Moreover, the CR$_s$ and the ILR$_p$ are not
coincident. This seems to indicate that the formation of retrograde
secondary bars is less constrained than for the direct ones. The
central decoupling is indeed made easier in counter-rotation than in
direct rotation, and does not require large amounts of gas.  Note that
the relative pattern speed between the two bars is similar for the
retrograde and the direct cases (2.4 and 2.1 respectively).

\begfig 12 cm
\nompsy{flos1.ps}
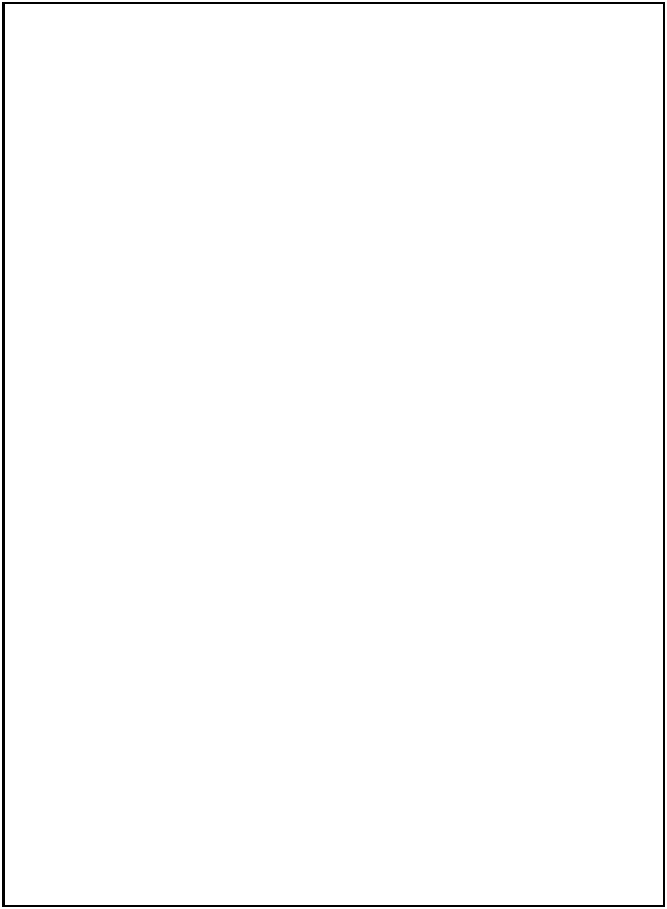
\figure{\FVELLOST a and b}{
Line-of-sight {\bf a} velocity $V_{\rm los}$, and {\bf b} velocity
dispersion $\sigma_{\rm los}$, for model $R_3$ at $t\!=\!1000$ (no
bars) of both components (crosses), at $t\!=\!2000$ ($\theta \approx
-60\degr$) of both components (filled circles), the direct component
(open circles), and the retrograde component up to 3~kpc (open
triangles).  The ``numerical slit'' is 0.4~kpc wide and is oriented
along the primary bar major axis.  Models are seen edge-on }
\endfig

\begfig 12 cm
\nompsy{flos2.ps}
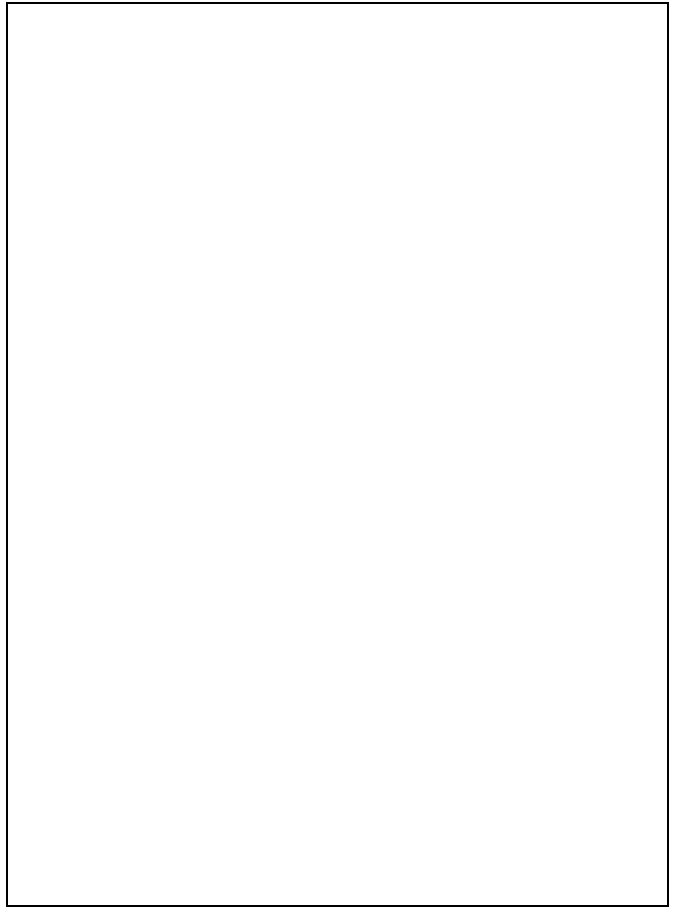
\figure{\FVELLOSC a and b}{
The same as Figs.~{\FVELLOST a and b} but for both components of model
$R_3$ at $t\!=\!2100$ ($\theta \approx 90\degr$) along the primary bar
major axis (filled circles), along the primary bar minor axis (open
circles), as well as for model $D_1$ at $t\!=\!1200$ ($\theta \approx
90\degr$) along the primary bar major axis (open triangles), and along
the primary bar minor axis (crosses).  Models are seen edge-on }
\endfig

\titlec{Kinematics}
Near the centre at $t\!=\!2000$, $V_t$ is still negative although much
less than at the beginning (see Figs.~{\FROTCUR a and b}). This is a
true counter-rotation.  Clearly, direct and retrograde double-barred
models have different $V_t$. Some rotation curves of SB0's along the
bar major axis show negative values near the centre (Bettoni 1989) but
it is not yet clear if this is only an apparent counter-rotation due
to projection effects on streaming motions or poor slit alignment with
the bar major axis (see Wozniak \& Pfenniger 1996 for a recent
discussion of this problem).

In order to be in a position to compare numerical models with
observations in a proper way, the line-of-sight velocity $V_{\rm los}$
and velocity dispersion $\sigma_{\rm los}$ have been computed through
a ``numerical slit''.  Owing to the particle noise, differences
depending on the slit alignment or $\theta$ are difficult to clearly
highlight, so that only the most favourable situation will be analysed
here, i.e. the edge-on case.

Model $R_3$ is displayed in Figs.~{\FVELLOST a and b} at two different
times, i.e. at $t\!=\!1000$ where no bar has yet developed, and at
$t\!=\!2000$ in the middle of the two counter-rotating bar phase.  In
the latter case, the slit is aligned along the primary bar, and
$\theta \approx -60\degr$.  These two models have at 0.5~kpc a
respective maximum net counter-rotation of $V_{\rm los} \approx 15 \,
\rm km \, s^{-1}$ and $V_{\rm los} \approx 35 \, \rm km \, s^{-1}$. 
The rotation of the individual components is also shown at
$t\!=\!2000$.  The formation of the primary bar results in a decrease
of $V_{\rm los}$ by a few tens of $\rm km \, s^{-1}$ beyond 2~kpc. The
most spectacular effect induced by the two bar formation is observed
on $\sigma_{\rm los}$; at $t\!=\!1000$ (as well as in the initial
conditions), its profile displays a central dip of $\approx 30 \, \rm
km \, s^{-1}$ which is almost fully erased by the secondary bar
formation at $t\!=\!2000$.  This model has a nearly flat profile up to
1.5~kpc ($\sigma_{\rm los} \approx 190 \, \rm km \, s^{-1}$),
i.e. close to $l_s$ seen at an angle of 60\degr. So, the central
galaxy part is considerably getting hotter with time.  An even deeper
dip has also been observed by Bertola \etal\ (1996) in NGC~3593. This
is a strong indication that this galaxy has not (yet?) formed a
central counter-rotating secondary or nuclear bar.  The central
$\sigma_{\rm los}$ of the retrograde component is $\approx 50 \, \rm
km \, s^{-1}$ smaller (i.e. colder) than the one of the direct
component.

Figures~{\FVELLOSC a and b} compare models $R_3$ at $t\!=\!2100$ and
$D_1$ at $t\!=\!1200$ with the slit i) aligned along the primary bar
major axis, and ii) aligned along the secondary bar major axis.  In
both models, the primary and secondary bar are nearly perpendicular
($\theta \approx 90\degr$).  In the model $R_3$, $V_{\rm los}$ shows a
maximum counter-rotation ($\approx 30 \, \rm km \, s^{-1}$) when the
slit is aligned along the secondary bar major axis, and in this case
$\sigma_{\rm los}$ displays a distinct peak of nearly $20 \, \rm km \,
s^{-1}$ above the otherwise observed flat profiles.  Although direct
and retrograde models are not fully comparable (gas is present in the
direct models), it is however instructive to highlight major
differences.  Models $R_3$ mainly differs from model $D_1$ by its true
central counter-rotation, and by its apparent rotation around 2~kpc
which is about $40 \, \rm km \, s^{-1}$ slower. The apparent
rotational support is thus only about $70 \, \rm km \, s^{-1}$. This
is much less than the circular velocity ($\approx 310 \, \rm km \,
s^{-1}$).  The two $\sigma_{\rm los}$ profiles are also very
different. The retrograde model has either a flat profile up to
1.5~kpc, or a steeper profile with a shoulder between 1.0 and 1.5~kpc.
Then $\sigma_{\rm los}$ decreases nearly linearly.  The velocity
dispersion of model $D_1$ is very similar at the centre ($\sigma_{\rm
los} \approx 185 \, \rm km \, s^{-1}$) but then steeply decreases. At
2~kpc, it is approximately $40 \, \rm km \, s^{-1}$ lower than the one
of model $R_3$.

In the face-on case, we have $\sigma_{\rm los} \!=\! \sigma_z$. With
time, the $\sigma_z$ radial profile is progressively becoming less
sharply falling.  Interestingly enough, the observed profiles of the
double-barred galaxies NGC~1291 and NGC~1543 by Jarvis \etal\ (1998)
have a very slow decrease as well!

\begfig 6 cm
\nompsy{fschema.ps}
\figure{\FSCHEMA}{
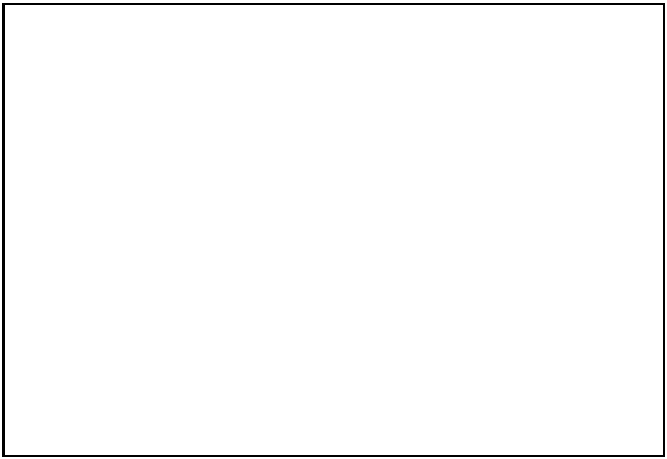
Schematic sketch of the effect of disc thickness and retrograde
angular momentum concentration on the galaxy stability and
evolution. The position of the models discussed in Sect.~4 are
indicated}
\endfig

In conclusion, the kinematics of counter-rotating bar-within-bar
models is clearly distinct (e.g. significant central counter-rotation
of $V_{\rm los}$; flat central $\sigma_{\rm los}$ profiles) from the
one of direct bar-within-bars models.  High resolution central
velocity field mappings of early-type galaxies should bring the
deciding proof concerning the secondary bar direction of
rotation. Moreover, in highly inclined galaxies with concentrated
counter-rotation (e.g. NGC~3593), one should easily distinguish
between disc and bar components from kinematic features
(i.e. truncated pyramid-like $\sigma_{\rm los}$ profile or
volcano-like $\sigma_{\rm los}$ profile).

\titlea{Discussion}
The dynamical evolution of disc systems with stellar counter-rotation
appears to be dependent on at least three parameters: The disc
thickness, the mass and the central concentration of the retrograde
component.  A schematic summary is displayed in Fig.~{\FSCHEMA}.
Although the whole space parameter have not yet been explored, the
presence of bending instabilities are clearly dependent on the disc
thinness (Sellwood \& Merritt 1994). Bending modes disappear in
thicker discs. In this case, the evolution is very different depending
on the concentration of retrograde angular momentum (which increases
when $a_1$ decreases and $M_1$ increases). A low concentration keeps
the system stable and axisymmetric.  An intermediate concentration
leads to the formation of a dynamically-dominant direct bar whose
retrograde phase-space entirely dominates the counter-rotating
component. A high concentration allows the dynamical decoupling
between the two components and leads to counter-rotating bars within
bars.

Interestingly enough, the best morphological similarities between
numerical models and typical SB0's occur in the models with
significant amount of central stellar counter-rotation. Indeed, if the
central part of the galaxy (the bulge) is co-rotating with the primary
bar, it also cooperates with the bar instability and generally aligns
with it unless i) it is hot enough, or ii) it significantly
counter-rotates (model $R_1$).  If the central mass is high enough, a
dynamical decoupling occurs both in the direct and retrograde cases
leading to the formation of bars within bars systems (models $D_1$ and
$R_3$). If the central mass is even higher, no bars can form.  For
instance, the presence of supermassive black holes in the nucleus
generates very hot centres and S0-like galaxies (Friedli 1994).  So, I
suspect that some lenticular galaxies should have recently accumulated
significant retrograde angular momentum, for instance by retrograde
satellite cannibalism.  Note that retrograde orbits are more stable
than direct ones, and thus the fraction of retrograde to direct
satellites should increase with time around galaxies.  The above
picture is consistent with the very high and convincing evidences for
an evolution along the Hubble sequence from late-types to early-types
(see e.g. Martinet 1995 and references therein). In particular, small
bulges (types Sc and Sb) can form from bars, bigger bulges (types Sa
and S0) can be generated by satellite accretion (Pfenniger 1993),
whereas elliptical galaxies can result from the merging of two or more
spirals (Barnes 1992).  Another interesting point is that the presence
of bars allows the progressive annihilation of the angular momentum of
each component in the regions with both direct- and counter-rotation.
This clearly leads to a decrease of the rotational support in favour
of pressure support allowing the formation of big bulges.  The central
spiraling of retrograde satellites should thus trigger the formation
of bigger bulges than direct ones. Distinct signs of counter-rotation
will thus disappear with time and discs within discs should survive
longer than bars within bars.

In Sect.~4.3, the viability of nested, counter-rotating, stable bars
has been demonstrated by numerical simulations. However, the initial
conditions are purely ad-hoc and certainly do not correspond to any
realistic formation scenario.  Indeed, the aim of this paper is not to
show a complete formation sequence but only to demonstrate that such
systems can exist, be long-lived, and to gain some insight on how they
evolve. Three more realistic formation schemes of such systems are
discussed below which will be investigated in more depth in a
subsequent paper.

\noindent
A. Fast accretion of a retrograde gas-poor satellite by an early-type
barred galaxy.  If a moderately compact and massive (5\%--15\% of the
host galaxy mass) gas-poor satellite on a retrograde orbit is
swallowed by a barred galaxy, dynamical friction will cause it quickly
to spiral to the centre and could produce there a counter-rotating
secondary bar.  The primary bar can already be formed or be induced by
the close interaction with the satellite.  No induced star formation
has to be considered.  There is however two major drawbacks: 1) If the
satellite is very compact, the merging of the galaxy and the satellite
should result in a significant dynamical heating which is able to
dissolve the primary bar and prevent the secondary bar from being
formed.  Moreover, observations of NGC~3593, NGC~4550 and NGC~7217
indicate cold, rotationally supported, galaxies.  Fast merging is
unlikely to have produced these systems.  In addition, other
signatures of the interaction (like $m\!=\!1$ modes, tails, ripples,
etc.) should be present as well.  2) On the contrary, if the satellite
is too loosely bound, it will be tidally disrupted and not enough mass
will remain in the centre to produce a distinct counter-rotating
component. So, a very fine tuning could be requested to produce in
this way two nested counter-rotating bars.

\noindent
B. Fast accretion of a retrograde gas-rich satellite by an early-type
barred galaxy. In this case, the evolution proceeds similarly to the
scenario A but at an expected faster rate due to significant
gas-dissipation.  The drawback 1) of scenario A remains.  The drawback
2) of scenario A is alleviated since even if the satellite is tidally
disrupted before reaching the centre, dissipation insures that the gas
will first settle into the disc plane and then spiral towards the
nucleus.  In particular, the gas coming from mass loss due to stellar
evolution in the direct component will interact with the retrograde
gas and strong dissipative shocks should occur.  Here, the main
problem concerns the way star formation is proceeding to transform the
central gas-rich component into a star-dominated one.

\noindent
C. Slow accretion of retrograde gas by an early-type barred galaxy.
The progressive, adiabatic, building up of a counter-rotating compact
gaseous disc by continuous or episodic infall is expected to preclude
excessive heating of the direct preexisting disc, removing thus the
drawback 1) of scenarios A and B.  Numerical simulations by Thakar \&
Ryden (1995) have shown that this is feasible but star formation is
not yet included in their models.  In the bar region, the gas should
be trapped by the stable retrograde $x_4$ and anomalous periodic orbit
families (inclined loops) which have a large extent. But again, for
the most part the gas has to be transformed into stars and it is not
yet clear if the counter-rotating component could concentrate enough
before becoming collisionless-dominated. However, different rates of
infall, initial retrograde angular momentum, star formation histories
could succeed in reproducing the various types of galaxies with
counter-rotation observed so far.

\titlea{Conclusions}
The main conclusions of this paper are the following:

\noindent
1) Self-consistent collisionless disc models of embedded
counter-rotating bars within bars can exist. Their lifetimes can at
least reach one Gyr.  Contrary to direct bars within bars models, no
gas is necessary.  Counter-rotating secondary bars may thus be
relevant for the gas-poor double-barred early-type galaxies
(e.g. NGC~1291).

\noindent
2) The presence of counter-rotating bars within bars produces peculiar
signatures in the central line-of-sight velocities of the numerical
models.  The observational existence of such systems in early-type
galaxies could thus be inferred through kinematic evidences.  In the
case of NGC~3593, the counter-rotating central component is apparently
not barred.

\noindent
3) The best morphological similarities between numerical models and
typical SB0's occur in the models with significant amount of stellar
nuclear counter-rotation. These galaxies might have either
adiabatically accumulated retrograde angular momentum to the centre or
recently swallowed a few retrograde satellites.

\noindent
4) In case of non-axisymmetric potential, the angular momentum of each
component is progressively transferred and cancelled out in the
regions with both direct- and counter-rotation.  So, with time the
signatures of the retrograde motion will disappear and the rotational
support will decrease in favour of pressure support allowing the
formation of big bulges.  The central spiraling of retrograde
satellites should trigger the formation of bigger bulges than direct
satellites.

\noindent
5) Self-consistent collisionless disc models of counter-rotating bars
within bars having the same mass, scale-length, axis ratio, and
pattern speed can also be constructed. However, thus far no connection
with any observed galaxies has been established.

\acknow{
This work has been supported by the University of Geneva (Geneva
Observatory) and the Swiss National Science Foundation (FNRS).  I
especially thank H.~Wozniak for having made available his ellipse
fitting code as well as L.~Martinet and D.~Pfenniger for a careful
reading of the manuscript, and the referee, J.H.~Hunter, for his
valuable comments. I wish also to thank C.~Davies for many fruitful
discussions.}

\begref{References}

\ref
Barnes J.E., 1992, ApJ 393, 484

\ref
Bender R., 1995, in: Galaxies in the Young Universe, Ed.~H.~Hippelein.
Springer, Heidelberg, (in press)

\ref
Bertola F., Cinzano P., Corsini E.M., \etal, 1996, ApJ, (in press)

\ref
Bettoni D., 1989, AJ 97, 79

\ref
Buta R., Crocker D.A., 1993, AJ 105, 1344

\ref
Ciri R., Bettoni D., Galletta G., 1995, Nat 375, 661

\ref
Combes F., 1994, in: Mass-Transfer Induced Activity in Galaxies,
ed.~I.~Shlosman. Cambridge University Press, Cambridge, p.~170

\ref
Davies C.L., Hunter J.H.~Jr., 1995, in: Waves in Astrophysics,
eds.~J.H.~Hunter~Jr., R.E.~Wilson.  New York Academy of Sciences, New
York, (in press)

\ref
Davies C.L., Hunter J.H.~Jr., 1996, MNRAS, (submitted)

\ref
de Vaucouleurs G., 1974, in: Formation of Galaxies, IAU Symp.~No~58,
ed.~J.R.~Shakeshaft. Reidel, Dordrecht, p.~335

\ref
de Vaucouleurs G., 1975, ApJS 29, 193

\ref
Friedli D., 1994, in: Mass-Transfer Induced Activity in Galaxies,
ed.~I.~Shlosman. Cambridge University Press, Cambridge, p.~268

\ref
Friedli D., 1996, in: Barred Galaxies, IAU Coll.~No.~157, eds.~R.~Buta
\etal. ASP Conference Series, (in press)

\ref
Friedli D., Benz W., 1993, A\&A 268, 65

\ref
Friedli D., Martinet L., 1993, A\&A 277, 27

\ref
Friedli D., Wozniak H., Rieke M., Martinet L., Bratschi P., 1996,
A\&AS, (submitted)

\ref
Galletta G., 1996, in: Barred Galaxies, IAU Coll.~No.~157, eds.~R.~Buta
\etal. ASP Conference Series, (in press)

\ref
Jarvis B., Dubath P., Martinet L., Bacon R., 1988, A\&AS 74, 513

\ref
Kuijken K., 1993, PASP 105, 1016

\ref
Levison H.F., Duncan M.J., Smith B.F., 1990, ApJ 363, 66

\ref
Martinet L., 1995, Fund. Cosmic Physics 15, 341

\ref
Merrifield M.R., Kuijken K., 1994, ApJ 432, 575

\ref
Pfenniger D., 1993, in: Galactic Bulges, IAU Symp.~No.~153,
eds.~H.~Dejonghe, H.J.~Habing. Reidel, Dordrecht, p.~387

\ref
Pfenniger D., Friedli D., 1993, A\&A 270, 561

\ref
Rix H.-W., Franx M., Fisher D., Illingworth G., 1992, ApJ 400, L5

\ref
Rubin V.C., Graham J.A., Kenney J.D.P., 1992, ApJ 394, L9

\ref
Sellwood J.A., Merritt D., 1994, ApJ 425, 530

\ref
Shaw M.A., Combes F., Axon D.J., Wright G.S., 1993, A\&A 273, 31

\ref
Shaw M.A., Axon D.J., Probst R., Gatley I., 1995, MNRAS 274, 369

\ref
Shlosman I., Frank J., Begelmann M.C., 1989, Nat 338, 45

\ref
Shlosman I., Begelmann M.C., Frank J., 1990, Nat 345, 679

\ref
Tagger M., Sygnet J.F., Athanassoula E., Pellat R., 1987, ApJ 318, L43

\ref
Thakar A.R., Ryden B.S., 1996, ApJ, (in press)

\ref
Wozniak H., Friedli D., Martinet L., Martin P., Bratschi P., 1995,
A\&AS 111, 115

\ref
Wozniak H., Pfenniger D., 1996, A\&A, (submitted)

\ref
Zang T.A., Hohl F., 1978, ApJ 226, 521

\endref
\bye